\documentclass{aa}  
\usepackage{amssymb}
\usepackage{starfont}
\usepackage[utf8]{inputenc}
\usepackage[draft]{hyperref}
\usepackage{graphicx}
\usepackage{txfonts}
\bibpunct{(}{)}{;}{a}{}{,}
\usepackage{color}

\begin{document} 

\titlerunning{Ground-based HCN submillimetre measurements in Titan's atmosphere}
   \title{Ground-based HCN submillimetre measurements in Titan's atmosphere: an intercomparison with Herschel observations\thanks{{\it Herschel} is an ESA space observatory with science instruments provided by European-led Principal Investigator consortia and with important participation from NASA.}}
   \author{M. Rengel\inst{1}
           \and
          D. Shulyak\inst{1,2}
          \and
          P. Hartogh\inst{1}
          \and
          H. Sagawa\inst{3}
          \and
          R. Moreno\inst{4}
          \and
          C. Jarchow\inst{1}
          \and
          D. Breitschwerdt\inst{5}
          }
   \institute{Max-Planck-Institut f\"ur Sonnensystemforschung,
              Justus-von-Liebig-Weg 3, 37077 G\"ottingen, Germany\\
              \email{rengel@mps.mpg.de}
    \and 
 Instituto de Astrof\'{\i}sica de Andaluc\'{\i}a - CSIC, c/ Glorieta de la Astronom\'{\i}a s/n, 18008 Granada, Spain
   \and           
Faculty of Science, Kyoto Sangyo University, Kyoto 603-8555, Japan
   \and        
LESIA -- Observatoire de Paris, CNRS, Universit\'e Paris 6, Universit\'e  Paris-Diderot, 5 place Jules Janssen, 92195 Meudon, France
   \and        
Zentrum f\"ur Astronomie und Astrophysik, Technische Universit\"at Berlin, Hardenbergstrasse 36, D-10623 Berlin, Germany 
          }          
    \date{Received May 31, 2021; accepted November 25, 2021}
  
    \abstract
   {}
   {The aim of this study is to measure the vertical distribution of HCN on Titan's stratosphere using ground-based submillimetre observations acquired quasi-simultaneously with the Herschel ones. This allows us to perform a consistency check between space and ground-based observations and to build a reference mean HCN vertical profile in Titan's stratosphere.}
   {Using APEX and IRAM 30-m, we obtained the spectral emission of HCN (4-3) and (3-2) lines.  Observations were reduced with GILDAS-CLASS. We applied a line-by-line radiative transfer code to calculate the synthetic spectra of HCN, and a retrieval algorithm based on optimal estimation to retrieve the temperature and HCN vertical distributions. 
We used the standard deviation-based metric to quantify the dispersion between the ground-based and Herschel HCN profiles and  the mean one.} 
   {Our derived HCN abundance profiles are consistent with an increase from 40\,ppb at $\sim$100\,km to 4\,ppm at $\sim$200\,km, which is an altitude region where the HCN signatures are sensitive.  We also demonstrate that the retrieved HCN distribution is sensitive to the data information and is restricted to Titan's stratosphere. The HCN obtained from APEX data is less accurate  than the one from IRAM data because of the poorer data quality, and covers a narrower altitude range. Comparisons between our results and the values from Herschel show similar abundance distributions, with maximum differences of 2.5 ppm ranging between 100 and 300\,km in the vertical range.
These comparisons also allow us to inter-validate both data sets and indicate reliable and consistent measurements. The inferred abundances are also consistent with the vertical distribution in previous observational studies, with the profiles from ALMA, Cassini/CIRS, and SMA (the latest ones below $\sim$230\,km). Our HCN profile is also comparable to photochemical models by \citet{2014Icar..236...83K} and \citet{2019Icar..324..120V} below 230\,km and consistent with that of \cite{2015Icar..247..218L} above 250\,km. However, it appears to show large differences with respect to the estimates by \citet{2015Icar..247..218L}, \citet{2018Icar..307..371D}, and \citet{2018ApJ...853...58L}  below 170\,km, and
by \citet{2018Icar..307..371D} and \citet{2018ApJ...853...58L} above 400\,km, although they are similar in shape. 
We conclude that these particular photochemical models need improvement.
\vspace{0.5cm}
}
   {}
   {}
   \maketitle
%
\section{Introduction}
The atmosphere of Titan, one of the moons of Saturn, is cold, dense, and nitrogen (N$_2$)-dominated, and exhibits a great diversity of molecules and a complex atmospheric chemistry. Hydrogen cyanide (HCN), a molecule crucial to the production of life's building blocks, is the main nitrile species observed in Titan's atmosphere, and indeed Titan has the most HCN-rich atmosphere in the Solar System. The detection of HCN in Titan's atmosphere is robust and its vertical profile has been determined by spectroscopic observations \citep{1991Icar...89..152C,1997Icar..126..170H,2002Icar..158..532M,2004ApJ...616L...7G,2007Icar..191..712V,2011A&A...536L...2C,2011aogs...25..173R,2014A&A...561A...4R,2016AJ....152...42M,2019Icar..319..417T,2019NatAs...3..614L}. HCN  is  generated  photochemically in Titan's atmosphere from reactions of hydrocarbon radicals with atomic nitrogen. The latter is produced from extreme ultraviolet (EUV) or electron impact on N$_2$, or is possibly liberated as a result of cometary impacts \citep{2011NatGe...4..359S}. HCN is produced at high altitudes, above 300\,km \citep{1996JGR...10123261L,2004JGRE..109.6002W} and removed by condensation deeper in the atmosphere, setting up a concentration gradient. A more recent  alternative  explanation proposed that HCN is thermodynamically generated via shock chemistry under lightning discharges in the low atmosphere \citep{2010Icar..207..938K}. 

HCN composition in Titan's stratosphere has been investigated based on a limited number of high-resolution submillimetre observations performed on June 23 and December 15, 2010, with the Herschel Space Observatory  \citep{2010A&A...518L...1P} using the Photodetector Array Camera and Spectrometer (PACS) \citep{2010A&A...518L...2P}, and  on July 16, 2010, using the Spectral and Photometric REceiver (SPIRE) \citep{2010A&A...518L...3G},  within  the  framework  of the guaranteed time key programme "Water  and  related chemistry  in  the  Solar  System"  (HssO)  \citep{2011A&A...532L...2H}. 
Measured HCN vertical distributions were consistent with an increase from 40\,ppb at $\sim$100\,km to $\sim$4 ppm at $\sim$200\,km, which is an altitude region where the HCN signatures are sensitive \citep{2011A&A...536L...2C,2014A&A...561A...4R}.
In  support of Herschel observations, we observed Titan from the ground at submillimetre (submm) and mm wavelengths using the 12-m single-dish Atacama Pathfinder Experiment (APEX) telescope located at 5100 m above sea level in the Atacama desert in northern Chile \citep{2006A&A...454L..13G} and with the Institut de Radioastronomie  Millimetrique  (IRAM)  30-m telescope in Granada, Spain. 
Comparing space-based observations with ground-based ones is important; a quantitative link between the inferred HCN abundances obtained by Herschel and ground-based observations is required to assess the quality of the data and to inter-validate them. The ground- and space-based observations were acquired in a time period corresponding to a very small fraction of a Titan year, and therefore we assume in the following analysis that temporal temperature variations are negligible. 
Here we report the ground-based observations and disk-averaged HCN measurements. Furthermore, the accuracy of the measurements is assessed through comparisons with previous, correlative results from Herschel and the literature, and we present a mean HCN profile obtained from our ground-based observations and the Herschel ones. 

Small planets (radius R $\leq$ 2 R$_\oplus$) are the most common in our Galaxy, and they continue to be discovered and characterised. Studies characterising Titan present an opportunity to investigate the atmospheric properties of analogous objects (Titan-like exoplanets) in order to understand their atmospheric characteristics. Here we also add a discussion about HCN in the atmospheres of exoplanets. Fiducial reference HCN abundances for atmospheric studies of Titan-like exoplanets are needed, and studies assessing whether or not  these data sets  are suitable for such purposes are essential.
   
\section{Observations and data reduction}

\subsection{APEX observations}
After having demonstrated the capabilities of APEX and of the APEX Swedish Heterodyne Facility Instrument (SHeFI APEX-1 receiver) for atmospheric observations on Titan \citep{2011aogs...25..173R}, HCN (4-3) at 345\,GHz was observed in Titan's atmosphere on June 16, 2010, at APEX\footnote{This publication is based on data acquired with the AtacamaPathfinder Experiment (APEX) under program ID 085.C-0910(A). APEX is a collaboration between the Max-Planck- Institut f\"ur Radioastronomie, the European Southern Observatory, and the Onsala Space Observatory.}. As the front end for the observations, we used the APEX-2 heterodyne receiver (SHeFI 345 GHz band; \cite{2008A&A...490.1157V}).
This receiver employs superconductor-insulator-superconductor (SIS) mixers and behaves as a single sideband receiver (SSB), providing a spectral resolution of 122\,kHz and  a  total  bandwidth of 1\,GHz.  The telescope was used in raster scan mode. Observing conditions were not optimal, which prevented us from acquiring the initially proposed observations with a signal-to-noise ratio (S/N) of 100 (5.5 h). Instead,  data were acquired with an on-source integration time of 31\,min, and an average S/N of only $\approx$8. Titan was observed near the western or eastern elongations at separation angles from Saturn of greater than 120\arcsec. Pointing and focusing of the telescope were regularly checked by scanning across Saturn in azimuth and in elevation (APEX has a pointing accuracy of 2\arcsec r.m.s. over the sky). 
The beam size of APEX at 352 GHz is 17.3\arcsec.  The apparent diameter of Titan was around 0.8\arcsec. 
\subsection{IRAM observations}
HCN (3-2) at 265.9\,GHz was observed on Titan with the IRAM 30-m and the Heterodyne Receiver Array (HERA) receiver on March 19, 2011\footnote{This work is based on observations carried out under project number  [145-10]  with  the  IRAM  30-m  telescope.  IRAM  is supported by INSU/CNRS (France), MPG (Germany) and IGN (Spain).}. The receiver also employs SIS mixers, and provides a spectral resolution of 4\,MHz and a total  bandwidth of 4\,GHz.  Observations were taken under good weather conditions ($\tau$  < 0.13; PWV < 2.5mm); the system temperature was 420\,K. The telescope was used in wobbler-switch mode.  The on-source integration time was 92\,min, allowing us to acquire a spectrum with S/N = 36. The beam size of IRAM-30m at 260 GHz is 9.5\arcsec. The apparent diameter of Titan was also around 0.8\arcsec.

\subsection{Data reduction}
The observations were reduced using the Continuum and Line Analysis Single-dish Software (CLASS) 
package of the Grenoble Astrophysics Group\footnote{\url{http://www.iram.fr/IRAMFR/GILDAS}}. 
CLASS follows standard data reduction processes for single-dish heterodyne spectroscopy; see for example \citet{pad}, \citet{p05}, and \citet{polehampton2013apex}.

\section{Radiative transfer modelling, retrieval of parameters, and results}

We computed the emerging radiance using a forward model described in \cite{1995SPIE.2586..196J}, \cite{j98}, and  \cite{hj}.  
This model was successfully applied to planetary spectra including those of Venus and Mars \citep{2008PSS...56.1368R,2010A&A...521L..48H}.  For Titan, the model consists of a line-by-line radiative transfer model that takes into consideration a homogeneous spherically symmetric atmosphere of Titan (grid of $127$ altitude points ranging from 0  to  1500\,km).  We integrated the intensity of outgoing radiation across the disk and limb of the planet to obtain total flux at each frequency.

Abundances of the main atmospheric molecules were adopted following \cite{2010JGRE..11512006N}:  0.984, 0.001, and 0.014 for N$_2$, H$_2$, and  CH$_4$, respectively. The main opacity sources at the frequencies of the HCN lines considered here are collision-induced absorption (CIA) due to N$_2$-N$_2$, which we took from \cite{1986ApJ...311.1043B}. We also checked the impact of other CIAs, in particular those due to N$_2-$CH$_4$ and N$_2-$H$_2$, and found them to be negligible. 
The transition parameters for both lines were taken from the 2016 edition of the high-resolution transmission molecular absorption (HITRAN) database \citep{2017JQSRT.203....3G}.

Regarding the vertical temperature--pressure (T-P) structure of Titan's atmosphere, we adopted the distribution used by \citet{2012Icar..221..753M} and \citet{2014A&A...561A...4R},  which is a  combination  of  the Huygens Atmospheric Structure Instrument (HASI) profile \citep{2005Natur.438..785F} below 140\,km, and the Cassini/Composite InfraRed Spectrometer (CIRS) stratospheric temperatures \citep{2010Icar..205..559V} above 140\,km.

For the initial vertical distribution of HCN, we adopted the result of \cite{2002Icar..158..532M}  obtained from millimetre observations at IRAM, which is a  well-probed reference distribution. Its use  offers  a  reliable  result in conjunction with  the  data  quality achieved with our observations, and  has  also  been  successfully applied  to Herschel/SPIRE and PACS observations \citep{2011A&A...536L...2C,2014A&A...561A...4R}, which also facilitates inter-comparisons.

The fitting of the APEX and IRAM 30-m spectra by the models and the retrieval of the temperature and HCN vertical profiles from the spectral data are achieved by successive iterations using an optimal estimation (OE) algorithm. The key idea of the OE algorithm is to retrieve the atmospheric state from the spectra by searching for the solution that provides an optimal balance between how well the model fits the data and the deviation of model parameters from their expected values. A detailed description of the OE algorithm is given by \cite{1976RvGSP..14..609R}. The OE algorithm implemented here is a \textsc{Python} package presented in \cite{2019AA629A109S}.

Neither of the HCN lines reaches a well-defined continuum level.  
The  temperature  scale  of  the  modelled spectra  was  in  units  of  flux density $S_\nu$. 
In order to compare the model with APEX observations, $S_\nu$ was converted  to antenna  temperature $T_a$ using $S_\nu$ = 24.4 $\times$ T$_a$ ·$\eta_f$ / $\eta_a$, where $\eta_f$ and $\eta_a$ are the forward and aperture efficiencies, respectively. In this study, we adopt the values considering the efficiencies listed on the 
APEX website\footnote{\url{http://www.apex-telescope.org/telescope/efficiency}}: $\eta_a$ is  60  and $\eta_f$ is 0.97  for 352\,GHz. These values are expected to have a 10\% uncertainty. In order to compare the model with IRAM observations, $S_\nu$ was converted  to antenna  temperature $T_a$. The relation between antenna temperature $T_a$ and flux density $S_\nu$ for IRAM is expressed as $S_\nu$=T$_a$/ $\Gamma$, where $\Gamma$ is the point source sensitivity of  the  antenna at 260\,GHz\footnote{\url{https://www.iram.es/IRAMES/mainWiki/Iram30mEfficiencies}}: 8.4. 

\subsection{Best-fitting solution} 

\begin{figure*}
   \centering
\includegraphics[width=18cm]{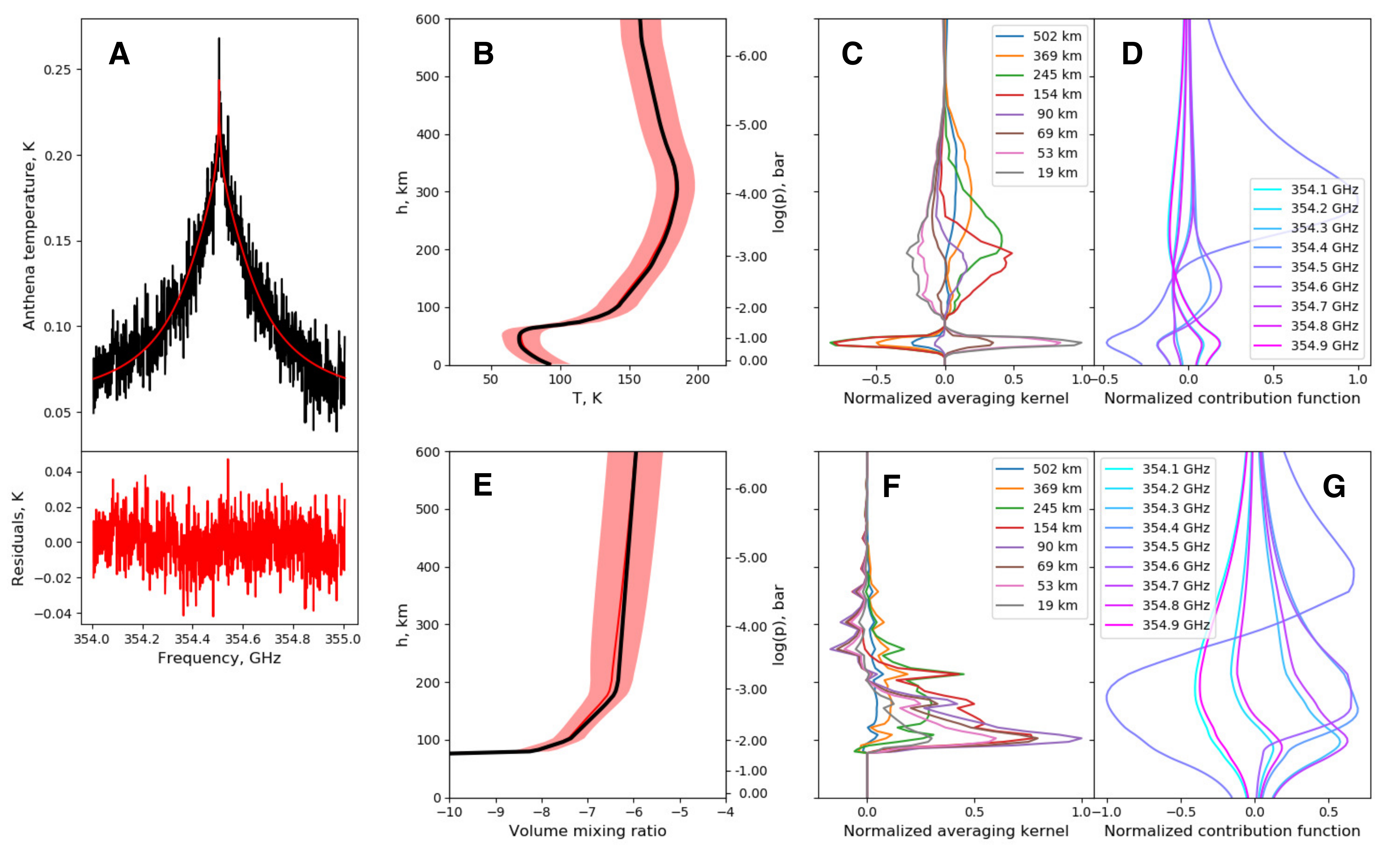}
 \caption{\textbf{A}: Comparison between observed and best-fit simulated HCN (4-3) lines (black and red,
        respectively, upper panel), and the difference between the observed and fitted spectra (lower panel). \textbf{B}: Retrieved temperature. \textbf{C}:  Corresponding averaging kernels. \textbf{D}: Corresponding normalised contribution functions. \textbf{E}: HCN distribution derived from the spectrum.  \textbf{F}: Corresponding averaging kernels. \textbf{G}: Corresponding normalised contribution functions. In B and E, the black and red lines show the initial and retrieved profiles, respectively, and the pink shadow shows the error bars. AVKs and contribution functions are shown for selected altitudes and frequencies, respectively, for better representation (see plot legends).}
              \label{fig1}%
    \end{figure*}

\begin{figure*}
   \centering
\includegraphics[width=18cm]{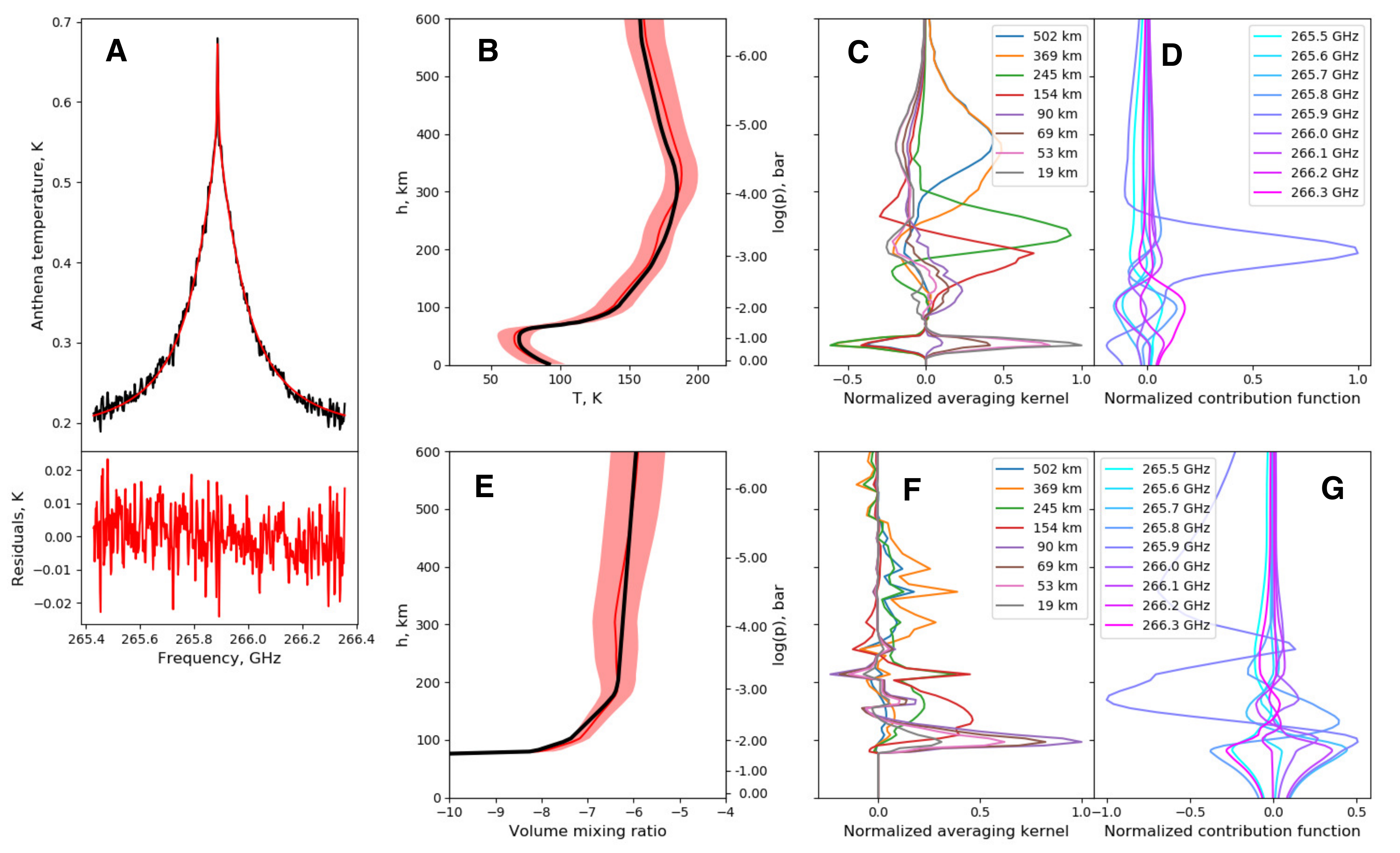}
 \caption{Same as Fig.~\ref{fig1}, but for the HCN (3-2) line.}
              \label{fig2}%
    \end{figure*}

Figures~\ref{fig1} and \ref{fig2} present the observations and show the best-fit between observed and modelled spectra, the difference between the two spectra (spectral residual), the temperature and HCN retrieval results,  the corresponding averaging kernel functions (AVKs) and contribution functions $D_y$ as defined by \citet{1990JGR....95.5587R}. Each AVK represents the sensitivity of the retrieval at a given altitude to variations in the true atmospheric state at all altitudes. Each contribution function shows how each channel contributes to the overall solution profile due to the measured intensity.
The vertical information content is given by the AVK functions where their amplitudes are different from zero. At these altitudes, our retrieval is sensitive to the true profile. 
Results suggest that our retrievals are sensitive in the vertical ranges of $\sim$50-480 km and 80-250 km for T and the mixing ratio for HCN (4-3), respectively, and  $\sim$50--550 km and 80--250 km for T and the mixing ratio for HCN (3-2), respectively.

The forward models (Figs.~\ref{fig1} and \ref{fig2}) show very good agreement with both observed HCN lines. Both instruments perform well  which allows us to constrain HCN abundance. The HCN abundance retrieved from IRAM data is more accurate than the one from APEX data.   
In particular, the APEX data can be satisfactorily fit with the model considering the profile of \citet{2002Icar..158..532M}. However, for IRAM data, we find that the retrieved HCN profile differs by a maximum factor of 2.5 from the reference profile  of \citet{2002Icar..158..532M} at altitudes of between 80 and 250~km, and the temperature differs by about 10~K from the adopted profile, both within the extent of the error bars.  We therefore conclude that our retrievals are consistent with the profile of \citet{2002Icar..158..532M}. Both lines allow us to retrieve limited altitude information range, meaning that the abundance of HCN cannot be constrained by the data below 80~km and above 250~km.

\subsection{Sensitivity of HCN retrievals to HCN \textit{a priori} }   
 
Retrievals of trace gas concentrations is a mathematically ill-posed problem and has non-unique solutions. There is potentially a family of solutions, with ones that are physically meaningful and others that are not, both being able to fit the spectrum equally well within the same error range. To check the reliability of the best-fit solution, we briefly investigate the sensitivity of the retrieval to the chosen \textit{a priori} solution.  If there is sufficient information available in a spectrum to constrain atmospheric properties, the result should not be affected by the \textit{a priori} atmospheric state. We retrieved HCN using three different \textit{a priori} profiles to show that the retrieved HCN converges to a reproducible profile in the altitude range covered by the measurements, that is, to a robust profile. 
Figures~3 and 4 show an example of this test, that is, thee results of retrievals assuming initial HCN constant profiles with altitude, with mixing ratios of 10$^{-8}$$\pm$10$^{-7}$, 10$^{-7}$$\pm$10$^{-6}$, and 10$^{-6}$$\pm$10$^{-5}$, respectively. 
We considered the reference temperature profile as described above with an associated uncertainty of 15~K. 

\begin{figure}
   \centering
\includegraphics[width=9cm]{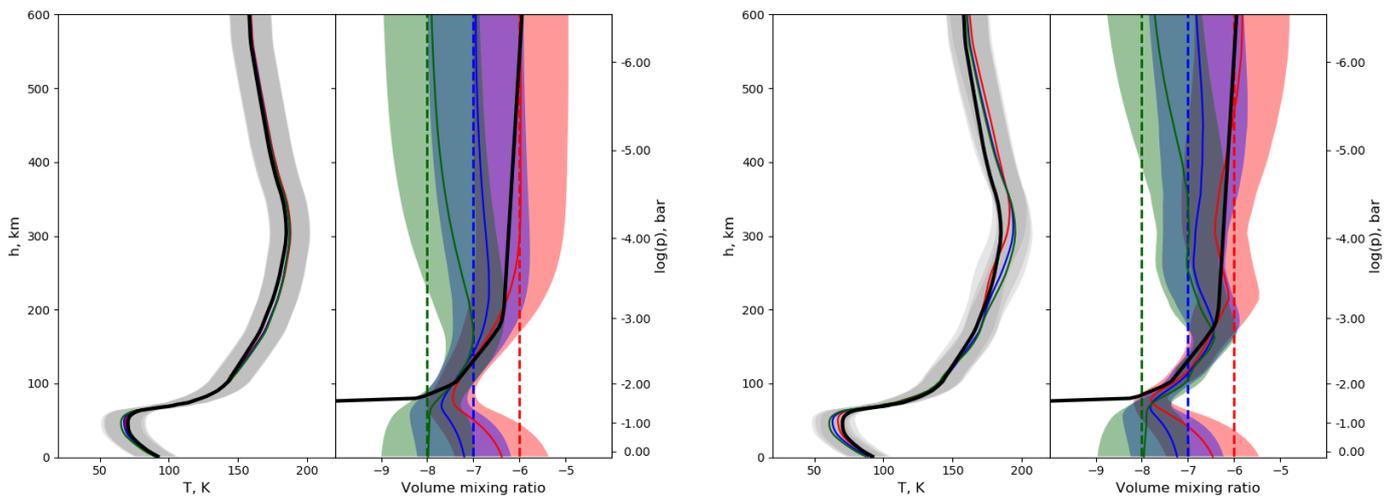}
 \caption{Comparisons of retrieved temperature and HCN profiles from a range of three different  \textit{a priori}  modified HCN profiles for the case of the APEX data. 
 Left:  Reference temperature shown in solid black, and retrieved temperatures in solid coloured lines. An associated temperature uncertainty at 15~K is shown by the shadowed area. Right: Different lines show \textit{a priori} HCN profiles (dashed) corresponding retrieved HCN profiles (solid) and their errors (shadowed area). The HCN profile of \cite{2002Icar..158..532M} is shown in solid black. }
              \label{fig3}%
    \end{figure}

For the IRAM observations (Fig.~\ref{fig4}), the retrieved HCN profiles share a common shape in the 80--180~km altitude region, where retrieval results are reported, demonstrating the validity of the HCN retrieval over this range. 
The retrievals from APEX data (Fig.~\ref{fig3}) are less accurate and are only robust in a narrower altitude range of 100--150~km compared to the IRAM case, mostly because of the rather poor S/N of the APEX data. 

The retrieved temperature profiles are very close to the initially assumed profile, with a maximum deviation of about 10~K in the case of IRAM data.  We also tried to consider different temperature profiles as our initial guess (e.g. isothermal ones) in order to assess the temperature sensitivity of the HCN lines. However, in all these cases, we failed to find a converged solution using APEX data. For the IRAM data, our retrieved temperatures were very different from the adopted profile, and HCN abundance was found to
vary drastically from one retrieval to another. Although these solutions provided fits to the observed HCN line of similar quality, we considered them non-physical and therefore excluded them from the current analysis; they are not shown here.

We show with this test that even considering the simplest possible HCN profile assumption, namely constant HCN profiles throughout the atmosphere, the measurements still allow us to
retrieve HCN profiles of similar shapes as the profiles derived with non-constant profiles.
All our calculations show that the derived HCN profiles are very close (within the error bars) to that provided by  \citet{2002Icar..158..532M}. 
We conclude here that the choice of the first guess has only a minimal effect on the retrieval results over the  levels  probed  by the measurements and that the data can constrain an unbiased  atmospheric  structure within that range.
A summary of the altitude heights probed by the measurements for T and mixing ratios and where HCN is robustly retrieved is given in Table\,1.

\begin{table*}
\caption{\label{t1}Summary of constrained altitude ranges}
\centering
\begin{tabular}{lccc}
\hline\hline
Data&T sensitivity range &HCN sensitivity range&Altitude range where \\
&[km]&[km]&HCN is robustly retrieved [km] \\
\hline
APEX HCN (4-3)&50--480 & 80--250&100--150\\
IRAM-30m HCN (3-2)&50--550& 80--250&80--180\\
\hline
\end{tabular}
\end{table*}

  \begin{figure}
   \centering
\includegraphics[width=9cm]{iram-multi-v2.pdf}
 \caption{Same as Fig.~\ref{fig3}, but for the case of the IRAM data.}
              \label{fig4}%
    \end{figure}
 
 \section{Discussion}    
\begin{figure*}
   \centering
\includegraphics[width=13cm]{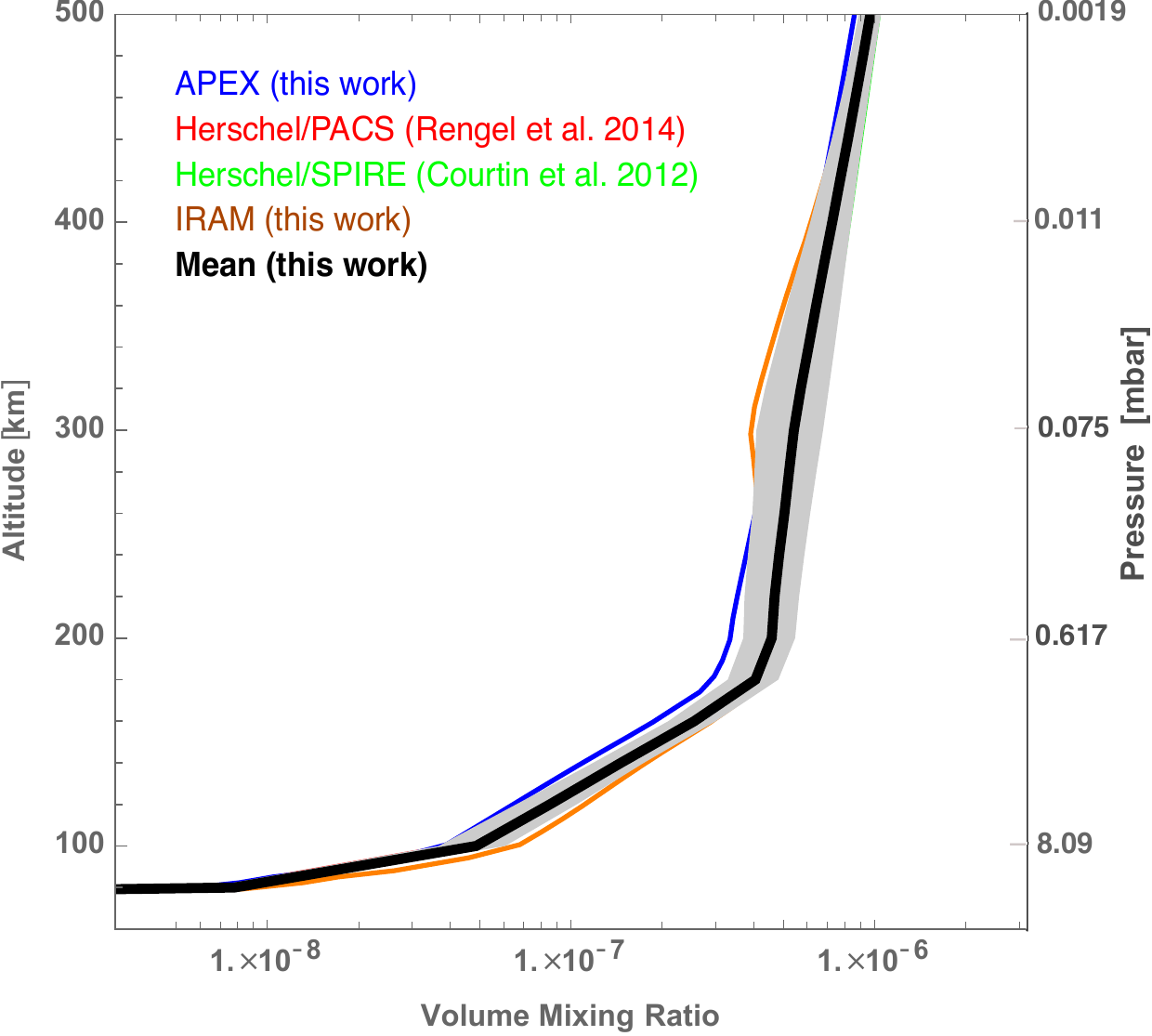}
 \caption{Vertical distributions of HCN obtained with APEX and IRAM 30-m (blue and yellow, respectively) compared with the Herschel profiles obtained by \cite{2011A&A...536L...2C} and \cite{2014A&A...561A...4R} (green and red, respectively). The black distribution shows the mean profile obtained from the four datasets, and the shaded region shows the associated 1-$\sigma$ standard deviation of the mean difference.}
              \label{fig5}%
    \end{figure*}

In Fig.~\ref{fig5} we present results of comparisons of HCN from APEX, IRAM30-m, and Herschel. In particular, we show the HCN mean profile obtained by considering four observations acquired quasi-simultaneously, and illustrate the associated 1-$\sigma$ standard deviation of the mean differences between the profiles. The HCN profiles derived here from the ground are in agreement with the findings from Herschel.
Our analysis confirms the result of \cite{2002Icar..158..532M}  from whole-disk mm observations. 
The four data sets show good agreement in shape and amounts of HCN, especially above 80\,km and below 250\,km. In this altitude range, there is a maximum difference of 2.5\,ppm in the  amount of HCN.

\begin{figure*}
   \centering
\includegraphics[width=17cm]{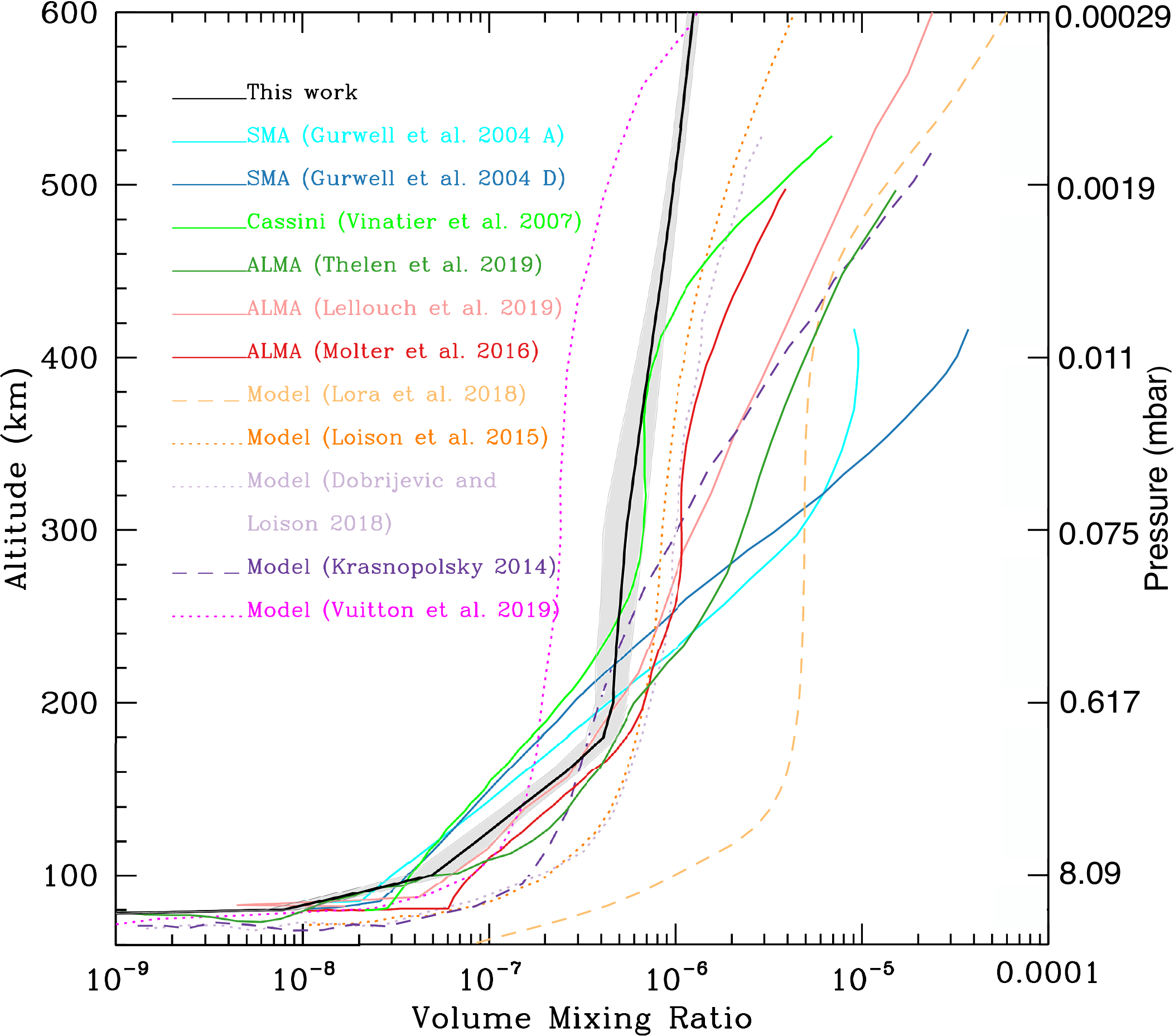}
 \caption{Mean HCN profile derived here (black) compared to observed profiles from the literature (coloured solid lines), and to predicted HCN profiles from photochemical models for Titan (coloured dashed lines) and for the atmosphere of planets around G stars \citep{2018ApJ...853...58L}. }
              \label{fig7}%
    \end{figure*}

The inferred abundances here are also consistent with the vertical distribution found in previous observational studies (Fig.~\ref{fig7}). The mean profile that we derive here is consistent with those derived from the Atacama Large Millimeter/submillimeter Array (ALMA) \citep{2019Icar..319..417T, 2016AJ....152...42M,2019NatAs...3..614L} and Cassini/CIRS, both from limb observations at 80$^{\circ}$ N \citep{2007Icar..191..712V} and from nadir observations inferred near the equator to altitudes of around 130\,km \citep{2007Icar..189...35C}. However, the abundance increase with altitude is less steep than in the Submillimeter Array (SMA)-derived profiles \citep{2004ApJ...616L...7G} above $\sim$230\,km. Above 250\,km, where our observations start to lose sensitivity,  we find that our HCN profile is consistent with the previous observations of \cite{2007Icar..191..712V} and the photochemical model of \cite{2015Icar..247..218L}.
Furthermore, our  mean HCN profile is also comparable to the photochemical models of  \citet{2014Icar..236...83K} and \citet{2019Icar..324..120V}  below 230\,km.  
The HCN modelled profiles by \citet{2015Icar..247..218L}, \citet{2018Icar..307..371D}, and \citet{2018ApJ...853...58L} appear to have over-predicted the amounts of HCN in atmospheres below 170\,km. However, the earlier ones are consistent in shape. These photochemistry models require revision, not only in the calculated absolute amount of HCN, but also in its vertical distribution below 170\,km and 250\,km, and, excluding \cite{2015Icar..247..218L}, above 400\,km as well. The model from \cite{2018ApJ...853...58L}  includes HCN modelled in the atmospheres of planets around G stars, with planetary parameters corresponding to Titan (more details in Section 5).

Figure~\ref{fig7} shows that measured HCN abundances on Titan with data acquired from space and the ground at similar epochs and with different transitions exhibit similar abundance distributions, and confirms that the former data set shows a small difference with respect to the ground-based  observations, with a difference that is essentially consistent and depends on the altitude level. This inter-validation allows us to derive reliable and consistent measurements. Beyond the intrinsic scientific interest, these observations prove their usefulness in supporting spacecraft observations of Solar System bodies, and in particular, of Titan's atmosphere.

We note that for the inter-comparisons presented here, we do not discuss the possible systematic effects due to different instruments and retrieval procedures. The impact of these latter in the comparisons are beyond the scope of this paper. 
Regarding possible temporal variability effects in Titan's atmosphere, the mean profile derived in this study confirms that the disk-averaged HCN does not vary significantly at these altitudes in Titan's atmosphere between 2010 and 2011 (our observations), 2012 and 2015 \citep{2019Icar..319..417T}, and in 2016 \citep{2019NatAs...3..614L}.  Furthermore, the disk-averaged temperature profiles of Titan obtained with ALMA were consistent within the error bars between 2012 and 2015 \citep{2018Icar..307..380T}, and were also consistent with the T--P profile used in this work,  justifying its adoption here.

\section{HCN in other planetary atmospheres}

Other planets can be seen as diverse possible laboratories for atmospheric and prebiotic chemistry. Here, we summarise the main findings from the literature for HCN in planetary atmospheres relevant to our study. In the Solar System, HCN is also present in the atmospheres of Pluto, Neptune, and Uranus, at concentrations of $\sim$40\,ppm, $\sim$1\,ppb, and 0.1\,ppb, respectively  \citep{1993ApJ...406..285M,2017Icar..286..289L}. HCN has been detected in Jupiter and upper limits of 0.93\,ppb have been placed \citep{1997Icar..130..387D}.
Cool terrestrial worlds with dense, hazy, and chemically complex atmospheres, that is, Titan-like exoplanets, could exist around a wide range of host stars. In those atmospheres, there may be different chemical pathways  leading HCN production and destruction, and these could be affected by variations in the far-to-near-ultraviolet ratio
(FUV/NUV ratio). While the main formation and loss pathways of HCN in Titan's atmosphere have been widely studied (\citet{2015Icar..247..218L}, \citet{2020ApJ...901..110P}, and references therein), little is known for Titan-like exoplanets.
Simulations of the atmospheric circulation and photochemistry of Titan-like exoplanets have been used to explore the sensitivity to host stellar type. 
It has been estimated that HCN mixing ratio profiles are similar between the different stellar spectra cases (G, K, and M stars) because HCN formation and loss are tied to the Lyman-$\alpha$ flux.
HCN abundances are slightly higher for the K dwarf case due to the higher N abundances from increased flux or photons \citep{2018ApJ...853...58L}. Modelled HCN in the stellar spectra case G holds constant planetary parameters at values corresponding to Titan, and these latter authors run their code with a default HCN profile provided by \cite{2007Icar..191..712V} and \cite{2015Icar..250..516L}. We find disagreement between our mean profile and the HCN-modelled profile from \cite{2018ApJ...853...58L}. The HCN modelled profile appears to have over-predicted the amounts of HCN in atmospheres of planets around a G star; these appear to be 100 times too large below $\sim$ 400\,km, even though the two profiles are consistent in shape below $\sim$ 400\,km.
The HCN-modelled profile from \cite{2018ApJ...853...58L} does not include the effects of condensation clouds, which are confined to the lower atmosphere.
The inferred HCN results obtained here could be assimilated as default profiles into climate models and chemistry calculations. 

Furthermore, simulations of the spectra of HCN in the IR have been found to display similar features considering the three different   stellar  cases  mentioned here \citep{2018ApJ...853...58L}. The response to variations of further climate-relevant parameters could be explored in order to further interpret  exoplanetary spectra, and to understand the key physical mechanisms shaping Titan-like exoplanetary atmospheres. Detailed simulations are beyond the scope of this paper.

HCN has been tentatively detected in the peculiar super-Earth 55 Cancri e with spectroscopic observations in the NIR \citep{2016ApJ...820...99T,2021arXiv210208965D} and in the hot Jupiter WASP-63b \citep{2017ApJ...850L..15M} with data obtained with the Hubble Space Telescope, and its detectability with future missions to observe super-Earths has been explored \citep{2019MNRAS.482.2893M}. HCN may also be present in ultra-hot Jupiters. In such atmospheres, a substantial degree of thermal ionisation and clouds may drive lightning and creation of HCN by ion-neutral chemistry. HCN has been searched for in hot H$_2$ atmospheres  with high-resolution spectroscopy, and an abundance of 10$^{-5}$  has been considered, placing a minimum limit on the HCN mixing ratio of log (HCN)=-6.5 in the atmosphere of the Hot Jupiter  HD~209458b \citep{2018ApJ...863L..11H}. Disequilibrium chemistry (vertical mixing) can enhance HCN abundances, dredging-up HCN to upper layers of the atmosphere and opening the possibility to detect it with future space-based facilities \citep{2020A&A...639A..48S}. HCN abundances in N$_2$-dominated atmospheres depend critically on the atmospheric C/O ratio, with significantly greater amounts of HCN generated photochemically when C/O $\geq$ 1 
\citep{2019Icar..329..124R}. Future data in the IR and dedicated space telescopes will help to shed more light on HCN in planetary atmospheres.

\section{Conclusions}

      We carried out complementary APEX and IRAM 30-m HCN (4-3) and (3-2) line observations, respectively, in Titan's atmosphere around the times of Herschel/PACS and SPIRE observations, and measured the HCN abundance using a retrieval algorithm based on optimal estimation. The quality and coverage of these data are sufficient for us to make a precise determination of the HCN abundance in the atmosphere of Titan at altitudes of 100-150 km and 80-180 km  from APEX and IRAM data, respectively. However, we note that the mixing ratio obtained from APEX data is less reliable than that derived from IRAM data because of the inferior data quality of the former.
      
Our main conclusions are as follows:       
      \begin{enumerate}
      \item We performed a consistency check and assessed the accuracy of the Herschel HCN observations by comparing them with ground-based observations. The HCN vertical profiles that we infer in this work are consistent with Herschel/PACS and SPIRE profiles, confirming the previous determination of  \cite{2002Icar..158..532M}.  Our retrieved HCN profiles are also consistent with the observed profiles from ALMA, Cassini/CIRS, and SMA (the latest ones below $\sim$230\,km). 
To the contrary, most HCN profiles that result from photochemical models display large deviations above 400\,km with respect to that retrieved here. 
   \item This study is relevant to the scientific community because the Herschel observations are publicly available\footnote{\url{http://archives.esac.esa.int/hsa/whsa/}} and may be further used in future studies of Titan's atmosphere. Our analysis shows that, with the current lack of space-based instruments observing Titan, the submm ground-based telescopes can successfully help to fill the consequent gaps in available data.
\item Here we show that our HCN profiles can be used as reference between 80 and 250\,km. 
For example, they could be used as input for modelling the 
atmospheres of hot super-Earths, as a guide to understanding what to expect in an N-dominated atmosphere, and as a reference in preparation for future observations of Titan and Titan-like exoplanets.
   \end{enumerate}
Observations of HCN with additional rotational lines --including rotational lines of isotopes-- in Titan's atmosphere and a search for HCN in N-dominated atmospheres are required to provide additional insight in order to improve models of Titan and Titan-like exoplanets

\begin{acknowledgements}
M.R. and D.S. acknowledge the support by the DFG priority program SPP 1992 "Exploring the Diversity of Extrasolar Planets (DFG PR 36 24602/41). D.S. acknowledges financial support from the State Agency for Research of the Spanish MCIU through the "Center of
Excellence Severo Ochoa" award to the Instituto de Astrof\'{i}sica de Andaluc\'{i}a (SEV-2017-0709).
We thank Juan Lora for providing profiles for Titan-like planets around diverse stars. We thank the anonymous  referee for the insightful comments and suggestions to the manuscript. We gratefully acknowledge the support of the APEX and IRAM staff during the observations. We acknowledge JPL's Horizons online ephemeris generator for providing Titan's position during the observations. This research has made use of NASA's Astrophysics Data System.

PACS has been developed by a consortium of institutes led by MPE (Germany) and including UVIE (Austria); KU Leuven, CSL, IMEC (Belgium); CEA, LAM (France); MPIA (Germany); INAF-IFSI/OAA/OAP/OAT, LENS, SISSA (Italy); IAC (Spain). This development has been supported by the funding agencies BMVIT (Austria), ESA-PRODEX (Belgium), CEA/CNES (France), DLR (Germany), ASI/INAF (Italy), and CICYT/MCYT (Spain).

SPIRE has been developed by a consortium of institutes led by Cardiff University (UK) and including Univ. Lethbridge (Canada); NAOC (China); CEA, LAM (France); IFSI, Univ. Padua (Italy); IAC (Spain); Stockholm Observatory (Sweden); Imperial College London, RAL, UCL-MSSL, UKATC, Univ. Sussex (UK); and Caltech, JPL, NHSC, Univ. Colorado (USA). This development has been supported by national funding agencies: CSA (Canada); NAOC (China); CEA, CNES, CNRS (France); ASI (Italy); MCINN (Spain); SNSB (Sweden); STFC, UKSA (UK); and NASA (USA).

\end{acknowledgements}

%
%
\bibliographystyle{aa}
\bibliography{biblio2}{}

\end{document}